\documentclass{emulateapj}
\usepackage{graphicx}
\usepackage{multirow}
\usepackage{color}

\defcitealias{cheng12a}{Paper~I}
\newcommand{\ngoodafe}{5620 }
\newcommand{\ngoodpm}{3985 }
\newcommand{\lag}{$\sim$15 km s$^{-1}$}
\newcommand{\lthin}{3.4}
\newcommand{\lthick}{1.8}
\newcommand{\lthinerr}{^{+2.8}_{-0.9}}
\newcommand{\lthickerr}{^{+2.1}_{-0.5}}

\newcommand{\rbracket}{]}

\begin{document}

\title{A Short Scale Length for the $\alpha$-Enhanced Thick Disk of the Milky Way: Evidence from Low-Latitude SEGUE Data}

\author{Judy Y. Cheng}
\affil{Department of Astronomy and Astrophysics, University of California Santa Cruz, Santa Cruz, CA 95064, USA}
\email{jyc@ucolick.org}

\author{Constance M. Rockosi\altaffilmark{1}}
\affil{UCO/Lick Observatory, Department of Astronomy and Astrophysics, University of California, Santa Cruz, CA 95064, USA}

\author{Heather L. Morrison}
\affil{Department of Astronomy, Case Western Reserve University, Cleveland, OH 44106, USA}

\author{Young Sun Lee}
\affil{Department of Physics and Astronomy and JINA: Joint Institute for Nuclear Astrophysics, Michigan State University, E. Lansing, MI 48824, USA}

\author{Timothy C. Beers}
\affil{National Optical Astronomy Observatory, Tucson, AZ 85719, USA}
\affil{Department of Physics and Astronomy and JINA: Joint Institute for Nuclear Astrophysics, Michigan State University, E. Lansing, MI 48824, USA}

\author{Dmitry Bizyaev}
\affil{Apache Point Observatory, Sunspot, NM, 88349, USA}

\author{Paul Harding}
\affil{Department of Astronomy, Case Western Reserve University, Cleveland, OH 44106, USA}

\author{Elena Malanushenko, Viktor Malanushenko, Daniel Oravetz, and Kaike Pan}
\affil{Apache Point Observatory, Sunspot, NM, 88349, USA}

\author{Katharine J. Schlesinger}
\affil{UCO/Lick Observatory, Department of Astronomy and Astrophysics, University of California, Santa Cruz, CA 95064, USA}
\affil{Department of Astronomy, The Ohio State University, Columbus, OH 43210, USA}

\author{Donald P. Schneider}
\affil{Department of Astronomy \& Astrophysics, Pennsylvania State University, University Park, PA 16802, USA}

\author{Audrey Simmons}
\affil{Apache Point Observatory, Sunspot, NM, 88349, USA}

\author{Benjamin A. Weaver}
\affil{Center for Cosmology and Particle Physics, New York University, New York, NY 10003 USA}

\altaffiltext{1}{Packard Fellow}

\begin{abstract}
We examine the $\alpha$-element abundance ratio, [$\alpha$/Fe], of \ngoodafe stars, observed by the Sloan Extension for Galactic Understanding and Exploration survey in the region 6 kpc $< R < 16$ kpc, 0.15 kpc $< |Z| < 1.5$ kpc, as a function of Galactocentric radius $R$ and distance from the Galactic plane $|Z|$. Our results show that the high-$\alpha$ thick disk population has a short scale length ($L_{\rm thick}\sim\lthick$~kpc) compared to the low-$\alpha$ population, which is typically associated with the thin disk. We find that the fraction of high-$\alpha$ stars in the inner disk increases at large $|Z|$, and that high-$\alpha$ stars lag in rotation compared to low-$\alpha$ stars. In contrast, the fraction of high-$\alpha$ stars in the outer disk is low at all $|Z|$, and high- and low-$\alpha$ stars have similar rotational velocities up to 1.5 kpc from the plane. We interpret these results to indicate that different processes were responsible for the high-$\alpha$ populations in the inner and outer disk. The high-$\alpha$ population in the inner disk has a short scale length and large scale height, consistent with a scenario in which the thick disk forms during an early gas-rich accretion phase. Stars far from the plane in the outer disk may have reached their current locations through heating by minor mergers. The lack of high-$\alpha$ stars at large $R$ and $|Z|$ also places strict constraints on the strength of radial migration via transient spiral structure.
\end{abstract}

\section{Introduction}\label{intro}
Detailed observations of the Galactic thick disk can be used to constrain the relative importance of cosmological accretion and secular processes in the formation and growth of the Milky Way disk. Thick disk stars are old ($>8$ Gyr, \citealt{ben05}) and provide a fossil record of the Galaxy at $z\sim2$. Observations of these old stars can serve as a complement to studies of distant galaxies at early times, many of which are seen to be thick, turbulent, clumpy, star-forming disks (e.g., \citealt{elm05,elm06,for09,for11}). Observations of nearby galaxies have shown that thick disks are common, with similar properties \citep{dal02,yoa05,yoa06,yoa08a,yoa08b}, which suggests that thick disks are a generic feature of disk galaxies. Thus, the processes responsible for the existence of the Milky Way thick disk may play an important role in the formation of all disk galaxies. 

Several mechanisms have been proposed to explain the formation of the thick disk. Four such mechanisms are: (1) vertical heating through minor mergers (e.g., \citealp{kaz08,rea08,vil08,kaz09,pur09,bir12}); (2) direct accretion of stars from satellites \citep{aba03}; (3) in-situ formation during an early turbulent disk phase due to high gas accretion rates (e.g., \citealp{bro04,bro05,bou09}); and (4) radial migration of stellar orbits via resonant interactions with transient spiral structure (e.g., \citealp{ros08a,ros08b,sch09a,sch09b,loe11}). Scenarios 1-3 fit within the context of hierarchical structure formation as predicted by $\Lambda$CDM cosmology, while scenario 4 is possible in a disk in complete isolation. Each of these scenarios can be tested through comparisons with the observed chemical and kinematic properties of stars in the Milky Way.

Since the thick disk's discovery by star counts \citep{yos82,gil83}, an apparent dichotomy between thin and thick disk populations has been established. Stars belonging to the thick disk are older and more metal poor (e.g, \citealt{gil95,chi00,ben04b,ive08}). In addition, thick disk stars have chemical abundance patterns distinct from the thin disk, with thick disk stars being enhanced in $\alpha$- and $r$-process elements \citep{edv93a,pro00,mas00,mas01,red03,red06,ben03,ben05,bre06}. The results of these latter studies show that the scatter in the observed [$\alpha$/Fe] is small, suggesting that thick disk stars formed quickly in a well-mixed interstellar medium. 

The high [$\alpha$/Fe] characteristic of thick disk stars indicates that they formed in a period of rapid star formation and chemical enrichment, during which Type II SNe were able to contribute significant amounts of $\alpha$-elements into the interstellar medium before Type Ia SNe increased the abundance of iron-peak elements. The observed abundance trends of nearby thick disk stars have been used to estimate that the thick disk formed over a period of $\sim$1-3 Gyr \citep{gra00,mas01,mas03,ben04b}. The chemical properties of thick disk stars are thus a powerful tool for understanding the chemical enrichment and star formation history of the Galaxy.

Whether the enhanced [$\alpha$/Fe] for thick disk stars in the solar annulus ($R_{\rm GC, \odot}=8.0$ kpc) is present for \textit{in-situ} thick disk stars (i.e., those at large $|Z|$, where the thick disk is expected to dominate) remains an open question. Early \textit{in-situ} studies examined only the [Fe/H] distribution \citep{gil95}. Analyses of more elements has largely been restricted to stars in the solar neighborhood, where stars are typically divided into thin and thick disk populations by their kinematics (e.g., \citealp{ben03,ben05}). Recently, \citet{ben11} examined the abundances of a sample of 119 red giants in the inner and outer disks using high-resolution spectroscopy and found evidence that the [$\alpha$/Fe] trends observed in the inner disk do not extend to the outer disk (see also \citealp{ben10b,alv10}). Thus, while many of the thick disk formation scenarios discussed above (such as the work of \citealt{bro05} and \citealt{sch09b}) are able to reproduce the dichotomy seen in the chemical properties of thin and thick disk stars in the solar neighborhood, observations have only begun to test the models at a wide range of $R$ and $|Z|$ in the Galactic disk. 

Using a sample of old disk stars from the Sloan Extension for Galactic Understanding and Exploration (SEGUE; \citealp{yan09}) survey, we have begun to explore a larger volume of the Galaxy. We previously showed that the radial metallicity gradient in [Fe/H] becomes flat for stars located at vertical heights $|Z|>1$ kpc from the Galactic plane, where the thick disk is expected to be the dominant population \citep[hereafter \citetalias{cheng12a}]{cheng12a}. This result is consistent with a chemically homogeneous thick disk, which is predicted by thick disk formation during a period of early gas-rich accretion (scenario 3). The flat gradient could also be explained if the strength of radial mixing is sufficient to erase a pre-existing gradient in the thin disk (scenario 4).

\citetalias{cheng12a} also demonstrated that the reported flattening trend in the radial metallicity gradient at $R\gtrsim10$ kpc \citep{yon05,luc06} could arise because all of the distant tracers were located at large $|Z|$, where the radial gradient is flat. We found that the observed trend could result from a simple superposition of a negative radial gradient at small $|Z|$ with a flat radial gradient at large $|Z|$. Because the clusters discussed in the literature at large $R$ were also located at large $|Z|$, it is unclear whether the observed trends are due to changes in the radial or vertical directions. Therefore, we stressed the importance of examining the trends in [Fe/H] as a function of both $R$ and $|Z|$.

In this work, we extend our analysis of abundance gradients in the Milky Way to examine the $\alpha$-element abundance ratio, [$\alpha$/Fe], as a function of both Galactocentric radius $R$ and distance from the plane $|Z|$, using a sample of \ngoodafe field stars from SEGUE. Our work is complementary to that of \citet{ben11}, as our sample of main sequence turnoff stars is more than an order of magnitude larger than their sample of 119 K giants. We use a subset of the sample from \citetalias{cheng12a}, which covers the region 6 kpc $< R < 16$ kpc, 0.15 kpc $< |Z| < 1.5$ kpc. We present our data and results in \S\ref{data} and \S\ref{results}, respectively. In \S\ref{measurescalelengths} we present estimates for the thin and thick disk scale lengths; the procedure and errors are described in more detail in the Appendix. We discuss the implications of our results in \S\ref{discussion}. 

\section{Data}\label{data}
Our sample consists of main sequence turnoff stars from the Sloan Extension for Galactic Understanding and Exploration (SEGUE, \citealt{yan09,eis11}), part of the Sloan Digital Sky Survey (SDSS; \citealt{yor00}). The data are obtained using the same CCD camera \citep{gun98}, telescope \citep{gun06}, and filter system \citep{fuk96} as the SDSS. In this paper, we use a subset of stars from the sample of 7605 main sequence turnoff stars in \citetalias{cheng12a}, which cover the region 6 kpc $< R < 16$ kpc, 0.15 kpc $< |Z| < 1.5$ kpc. Briefly, these stars are selected using a cut in $g-r$ color. Stellar parameters $T_{\rm eff}$, log $g$, [Fe/H], and [$\alpha$/Fe] are determined from low resolution ($R\sim2000$) spectra using the SEGUE Stellar Parameter Pipeline (SSPP, \citealt{lee08a,lee08b,all08,smo11,lee11a}). In the present work, we select the stars with sufficient signal-to-noise ($S/N > 20$ pixel$^{-1}$, where each pixel corresponds to $\sim1{\rm\AA}$) for good [$\alpha$/Fe] measurements, which yields 5771 stars with [$\alpha$/Fe] measured to a precision of 0.1 dex \citep{lee11a}.

The $S/N$ cut effectively imposes a magnitude limit, and because the [$\alpha$/Fe] of a star will not affect its magnitude significantly, this magnitude limit does not bias our sample against low-$\alpha$ stars. Dartmouth isochrones \citep{dot08} show that a 0.2 dex difference in [$\alpha$/Fe] is equivalent to $< 0.1$ mag difference in $g$-band magnitude, which corresponds to a $10\%$ error in distance. This is a small effect compared to the total error in distance $20\%-25\%$ estimated in \citetalias{cheng12a}. Therefore, we do not expect any significant systematic differences between the volumes sampled by high- and low-$\alpha$ stars. In our sample, the high- and low-$\alpha$ stars span the same ranges in $R$ and $|Z|$, and they are seen out to the same distances.

As described in \citetalias{cheng12a}, we assign a weight to each target to correct for selection effects. The weight accounts for three properties of the selection: (1) Objects in regions with the highest extinction in each line of sight were not considered for spectroscopy. (2) Not all candidates for spectroscopy are observed. (3) The $g-r$ color cut introduces a bias against redder, metal-rich stars. We show in \citetalias{cheng12a} that using the calculated weights successfully reproduces the true metallicity gradients in a mock catalog (\S6.3.2). For a detailed discussion of the selection biases and how we correct for them using our weighting scheme, see \S4 and the Appendix of \citetalias{cheng12a}. 

In addition to the weights we calculated in \citetalias{cheng12a}, we apply a weight to account for the stars that are removed by the additional $S/N$ cut imposed on this sample; this is a small effect compared to the other weights. Taking all stars with non-zero weights, we are left with a sample of \ngoodafe main sequence turnoff stars. Most of the stars that are given weights of zero are very blue objects, which are likely to be hotter stars that are not on the main sequence (for more discussion see the Appendix of \citetalias{cheng12a}). 

Distances were determined using photometric parallax methods, by comparing the SSPP stellar parameters and photometry to the theoretical isochrones of \citet{an09}, as described in \citetalias{cheng12a}. Using a mock catalog of stars generated from the model of \citet{sch09a}, we estimate the errors in the distances to be $\sim20\%-25\%$; see \S6 of \citetalias{cheng12a} for details.

\section{Results}\label{results}
\subsection{Abundance Trends as a Function of $R$ and $|Z|$}\label{abundances}
\begin{figure*}[!t]
\epsscale{1.0}
\plotone{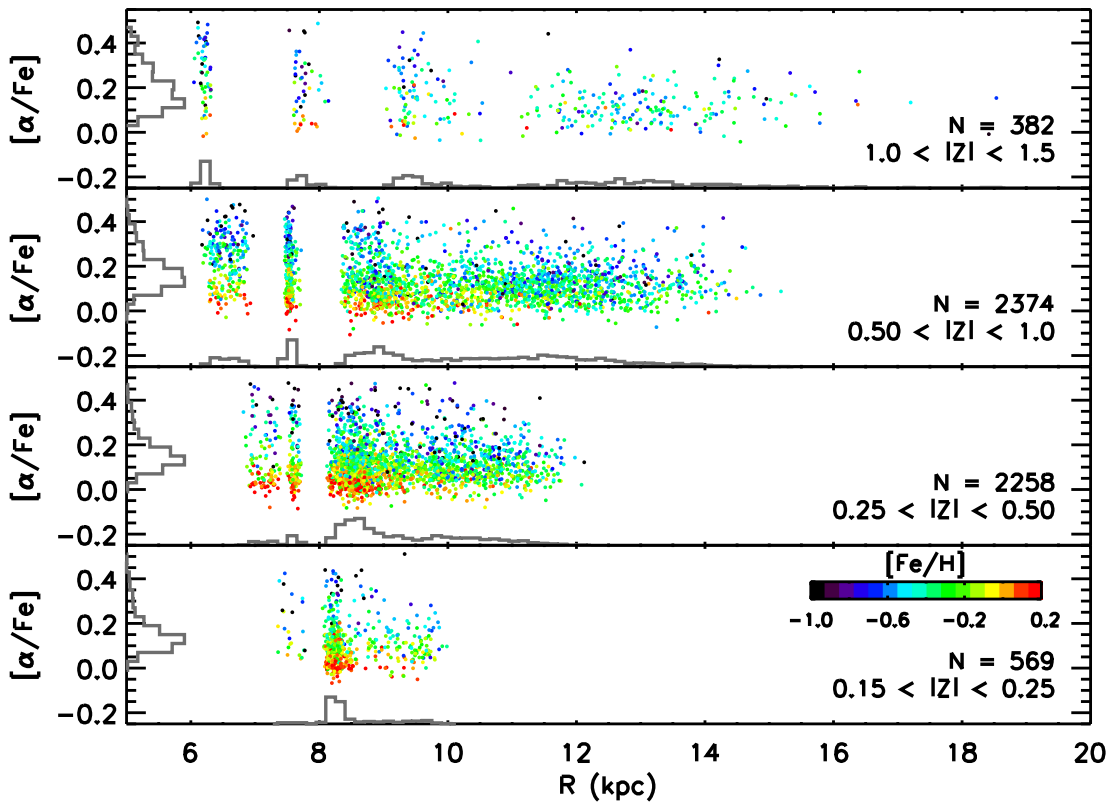}
\caption[$\alpha$-element abundance ratio [$\alpha$/Fe\rbracket~vs. Galactocentric radius $R$ in four $|Z|$ slices.]{$\alpha$-element abundance ratio [$\alpha$/Fe] vs. Galactocentric radius $R$ in four $|Z|$ slices. The SEGUE data are shown as dots, colored coded by [Fe/H]. At all $|Z|$, the majority of the high-$\alpha$ stars are located at small $R$ ($< 10$ kpc).}
\label{radial_zbin}
\end{figure*}

Figure~\ref{radial_zbin} shows [$\alpha$/Fe] as a function of Galactocentric radius $R$, in four slices of $|Z|$, for our sample of main sequence turnoff stars, color coded by [Fe/H]. Most of the high-$\alpha$ population is confined to small radii ($R<10$ kpc), consistent with the results of \citet{ben11}, who found a lack of high-$\alpha$ stars in the outer disk. In our sample, this lack of high-$\alpha$ stars is seen at all $|Z|$.

\begin{figure*}[!t]
\epsscale{1.0}
\plotone{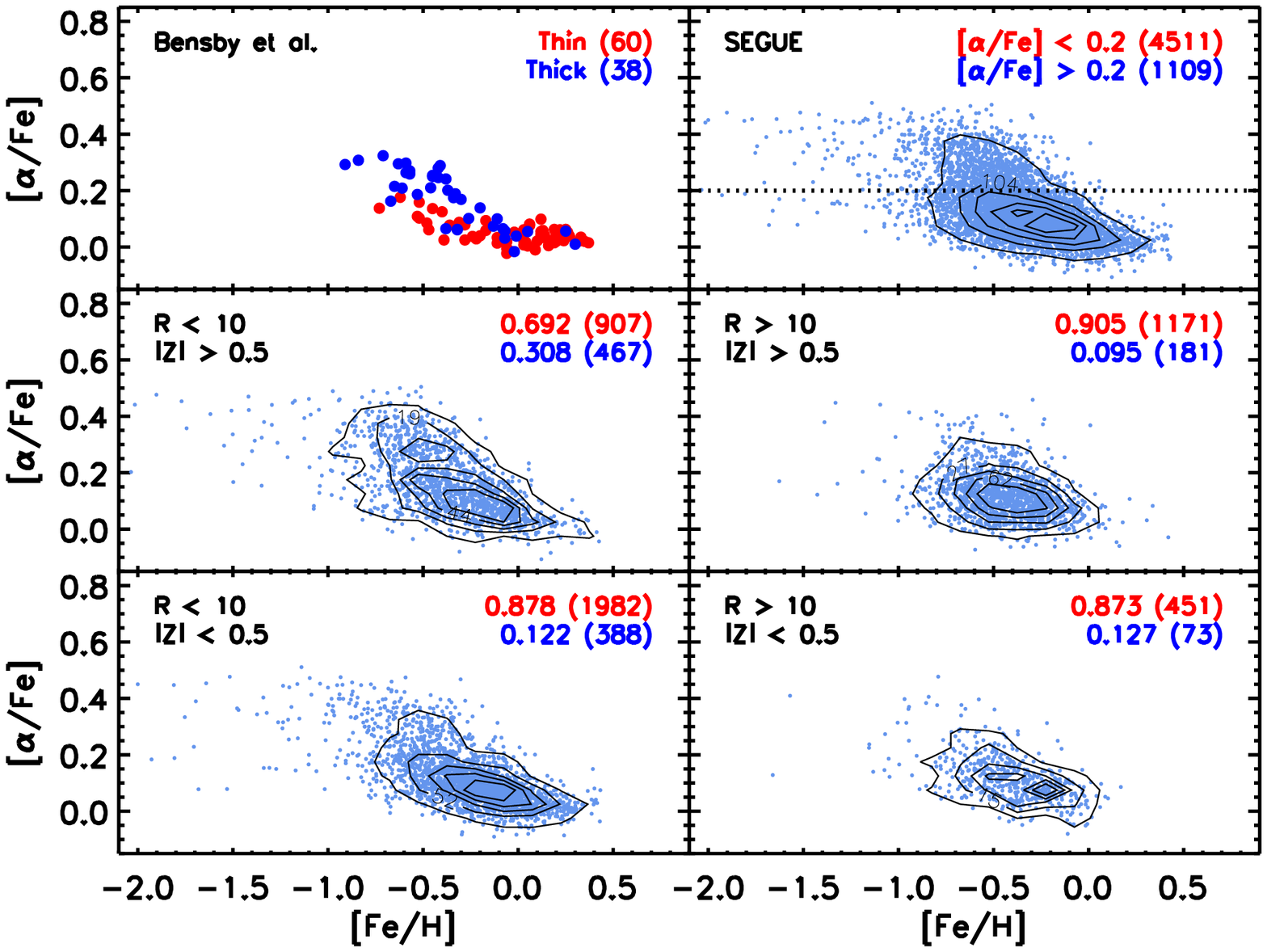}
\caption[Abundance trends [$\alpha$/Fe\rbracket vs. [Fe/H\rbracket.]{Abundance trends [$\alpha$/Fe] vs. [Fe/H]. Top left panel: the solar neighborhood sample of \citet{ben03,ben05}, with thin and thick disk stars (red and blue, respectively) assigned according to their kinematics. Top right panel: the total SEGUE sample. The horizontal dotted line indicates where we make the distinction between low- and high-$\alpha$ stars. Bottom four panels: the SEGUE sample, divided into four bins of $R$ and $|Z|$. The labels on the contours indicate the number of objects in a box with dimensions of 0.15 dex in [Fe/H] and 0.05 dex in [$\alpha$/Fe]. In each panel, the weighted fraction of high- and low-$\alpha$ stars (blue and red, respectively) is indicated, with the raw number of stars in each population in parentheses. The abundance patterns at $R < 10$ kpc are similar to those seen for thin and thick disk stars in the solar neighborhood, with the fraction of high-$\alpha$ stars increasing at large $|Z|$. At $R > 10$ kpc, the fraction of high-$\alpha$ stars is low at \textit{all} $|Z|$.}
\label{hist_afe}
\end{figure*}

Figure~\ref{hist_afe} shows abundance trends in [Fe/H] and [$\alpha$/Fe]. The top left panel shows the solar neighborhood sample of \citet{ben03,ben05}, in which stars are assigned to the thin and thick disks (red and blue, respectively) according to their kinematics. In the solar neighborhood, kinematically hot stars (i.e., thick disk stars) are $\alpha$-enhanced relative to kinematically cooler stars (i.e., thin disk stars) at the same [Fe/H]. The top right panel shows the total SEGUE sample, where we see two populations analogous to the solar neighborhood thin and thick disk stars: (1) low-$\alpha$ stars that, like solar neighborhood thin disk stars, appear to follow a linear trend, with [$\alpha$/Fe] slightly decreasing as [Fe/H] increases, and (2) a tail of high-$\alpha$ stars that, like solar neighborhood thick disk stars, are more metal poor than the low-$\alpha$ population. 

In the remainder of our analysis, we divide our sample into high- and low-$\alpha$ stars, with the goal of comparing the high-$\alpha$ (low-$\alpha$) stars to the kinematically selected thick (thin) disk stars in the Bensby sample; this is similar to the chemical separation done by \citet{lee11b}. We make the cut at [$\alpha$/Fe]$=+0.2$, where the number of stars appears to drop dramatically at large $R$, as seen in Figure~\ref{radial_zbin}. This is marked by the horizontal dotted line in the top right panel of Figure~\ref{hist_afe}. The number of high- and low-$\alpha$ stars is indicated in the top right corner in parentheses.

Finally, in the bottom four panels, we show the SEGUE sample divided into four $R, |Z|$ bins with cuts at $R=10$ and $|Z|=0.5$ kpc. The numbers in the top right corner of each panel indicate the weighted fractions of the high- and low-$\alpha$ populations, with the raw number of stars in parentheses. The weighting has the effect of slightly decreasing the fraction of high-$\alpha$ stars in each bin, but the effect is not dramatic. This is likely because selection on $g-r$ color is biased against metal-rich stars (see \citetalias{cheng12a} for details); metal-poor stars, which are more likely to be high-$\alpha$ stars, are weighted less heavily to compensate for the bias.

At small $R$ ($<10$ kpc, left panels), we see the same high- and low-$\alpha$ populations as in the total SEGUE sample. The presence of two populations is especially evident at $|Z| > 0.5$ kpc. The fraction of high-$\alpha$ stars increases toward large $|Z|$ ($>0.5$ kpc), from $12\%$ to $31\%$, which is what we expect if the contribution from a high-$\alpha$ thick disk is greater far from the plane. At large $R$ ($>10$ kpc, right panels), in contrast to what is seen at small $R$, the fraction of high-$\alpha$ stars does not increase at large $|Z|$ ($>0.5$ kpc); the fraction is low at all $|Z|$. This observation at large $R$ is inconsistent with the picture of a high-$\alpha$ population associated with a thick disk component, which should become more dominant at large $|Z|$ at all $R$. The main result of Figure~\ref{hist_afe} is that the [Fe/H]-[$\alpha$/Fe] properties for stars at small $R$ are consistent with those found for solar neighborhood stars, while at large $R$ there is a lack of high-$\alpha$ stars. Furthermore, the fraction of high-$\alpha$ stars at large $R$ does not increase with $|Z|$, as expected if there is a high-$\alpha$, thick disk population in the outer disk.

\begin{figure*}[!t]
\epsscale{1.0}
\plotone{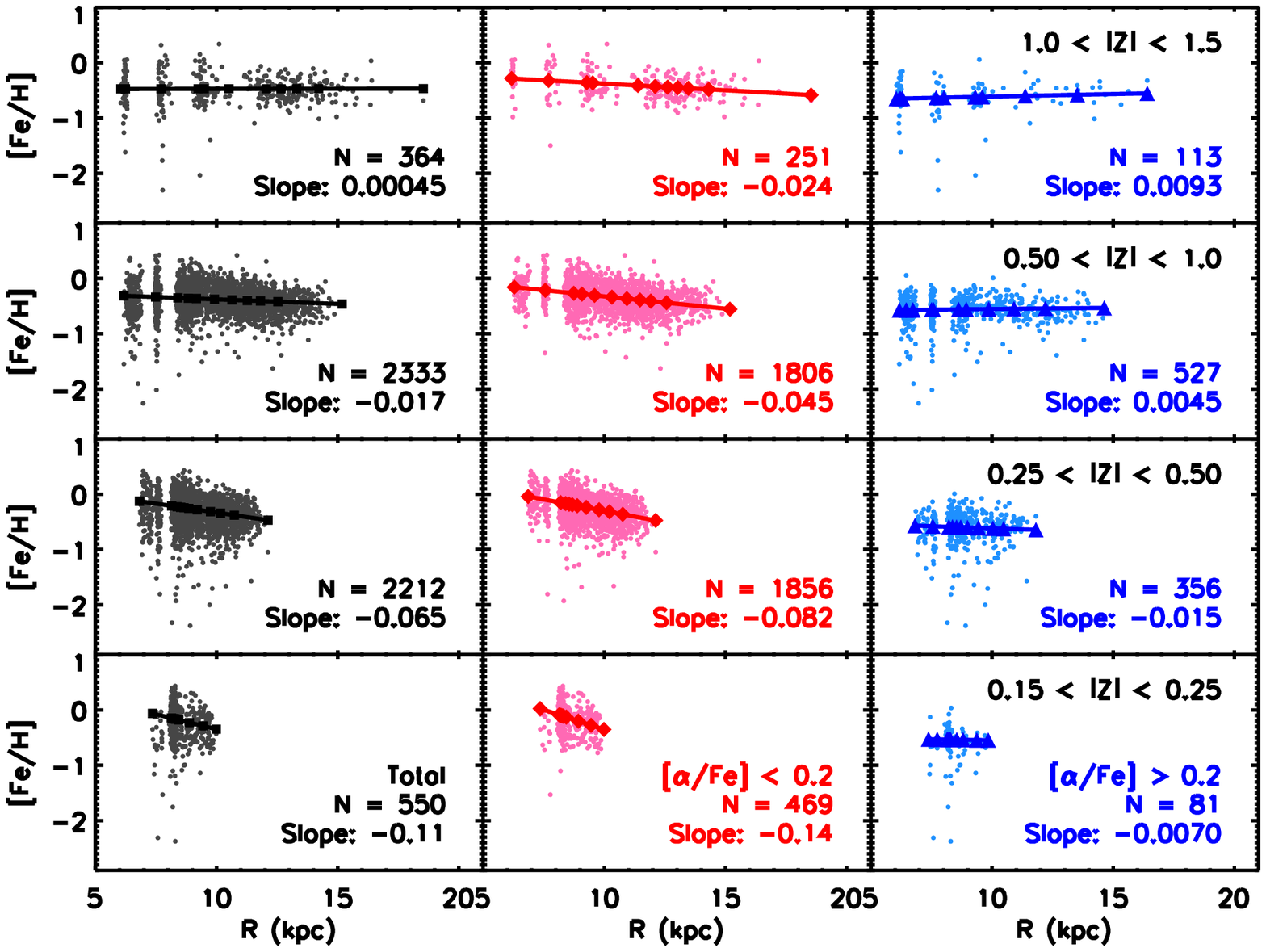}
\caption[Galactocentric radius $R$ vs. [Fe/H\rbracket in four $|Z|$ slices for high- and low-$\alpha$ stars.]{Galactocentric radius $R$ vs. [Fe/H] in four $|Z|$ slices for high- and low-$\alpha$ stars. Left column: Total sample. Middle column: Low-$\alpha$ stars ([$\alpha$/Fe] $< +0.2$). Right column: High-$\alpha$ stars ([$\alpha$/Fe] $> +0.2$). In each panel the raw number of stars and the measured slope are indicated in the bottom right corner. The lines show a linear fit to the data, with each star weighted to account for selection biases. The spacing of the symbols on the linear relation indicates the radial distribution of the stars. The radial gradient in the total and low-$\alpha$ samples become flatter at large $|Z|$, while the radial metallicity gradient of the high-$\alpha$ sample is flat at all $|Z|$.}
\label{alphas_rfeh}
\end{figure*}

The change in fraction of high-$\alpha$ stars at large $R$ suggests that the chemical abundances of stars change with Galactocentric radius, even at large $|Z|$. But how can this result be reconciled with the flat radial metallicity gradient at $|Z|>1.0$ kpc that we found in \citetalias{cheng12a}? Figure~\ref{alphas_rfeh} shows the radial metallicity gradient $\Delta$[Fe/H]/$\Delta R$ in four $|Z|$ slices for the total sample (gray, left column) and divided into low-$\alpha$ (pink, middle column) and high-$\alpha$ stars (blue, right column). For each sample, we fit a linear trend to the data, with each point weighted to account for selection biases, as in \citetalias{cheng12a}. The number of stars and the slope of a linear fit to the data are indicated in the bottom right corner of each panel. 

\begin{figure}[!t]
\epsscale{1.0}
\plotone{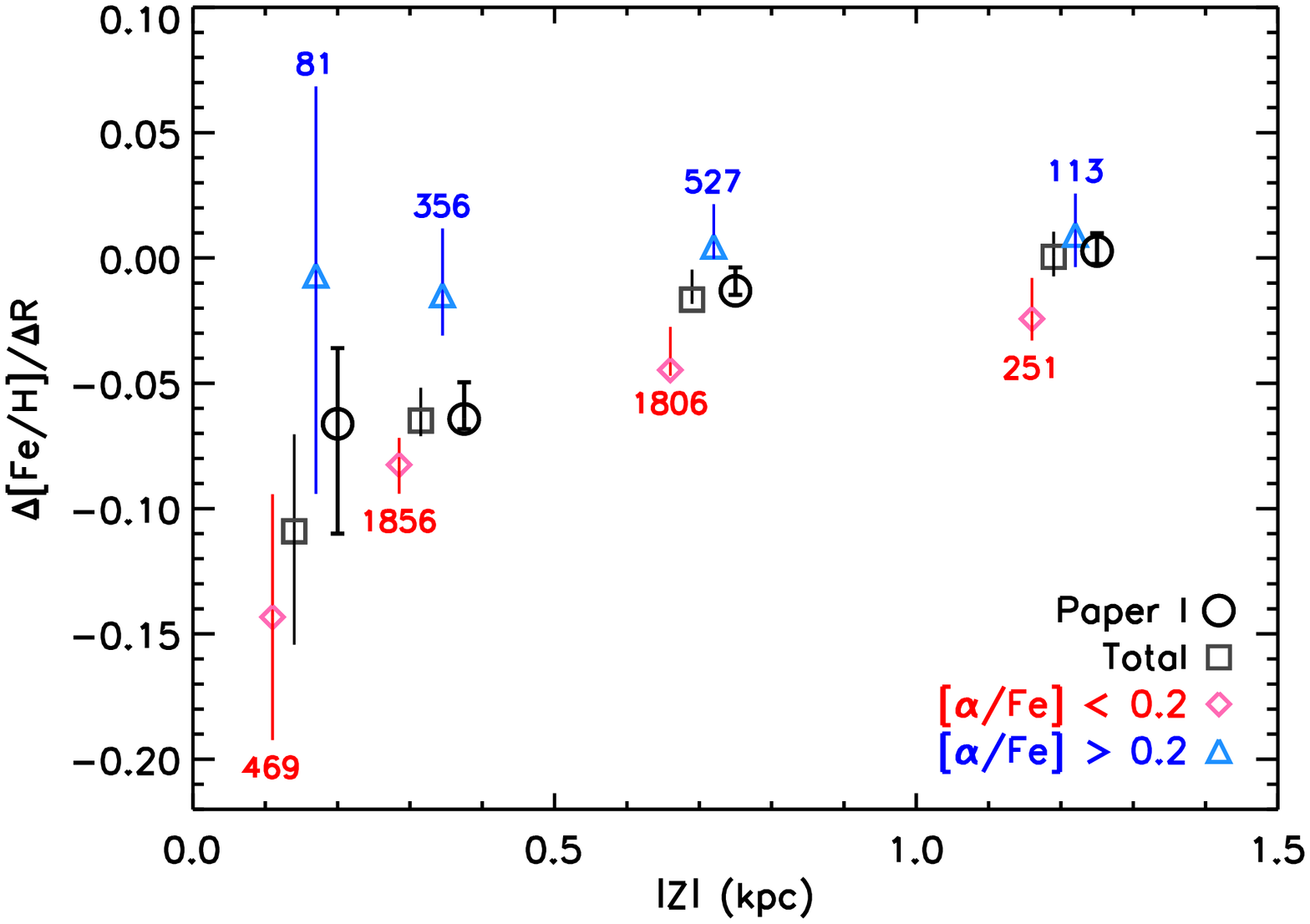}
\caption[Radial metallicity gradient, $\Delta$[Fe/H\rbracket/$\Delta R$, vs. distance from the plane, $|Z|$, for high- and low-$\alpha$ stars.]{Radial metallicity gradient, $\Delta$[Fe/H]/$\Delta R$, vs. distance from the plane, $|Z|$, for high- and low-$\alpha$ stars. The radial gradients for the total sample in this work (gray squares) are consistent with those of \citetalias{cheng12a}, which were measured using a larger sample with a less strict $S/N$ cut (black circles). The change in the radial gradient of the total sample as a function of $|Z|$ is driven by the change in the radial gradient of the low-$\alpha$ stars (pink diamonds); the radial gradient of the high-$\alpha$ stars (blue triangles) shows no obvious trend and is consistent with a flat gradient (slope of zero) at all $|Z|$. The number of stars used in each gradient measurement in the low- and high-$\alpha$ samples is indicated. The error bars reflect the random errors in the gradient measurement due to errors in the stellar parameters (see \S6 of \citetalias{cheng12a} for details).}
\label{alphas_slopes}
\end{figure}

The change in $\Delta$[Fe/H]/$\Delta R$ with $|Z|$, as shown in Figure~\ref{alphas_rfeh}, is summarized in Figure~\ref{alphas_slopes}. The radial gradient of the high-$\alpha$ sample (blue triangles) is flat at all $R$ and $|Z|$, but it is not solely responsible for the flattening trend with $|Z|$ seen in the total sample (gray squares). The high-$\alpha$ stars do, however, make the gradients for the total sample flatter, especially at $|Z|>1.0$ kpc, where the fraction of high-$\alpha$ stars is the largest. The flattening trend of the low-$\alpha$ sample (pink diamonds) is closely followed by the trend in the total sample. The results of \citetalias{cheng12a} are also shown (black circles) in Figure~\ref{alphas_slopes}. These are slightly different than the gradients measured for the total sample because of the $S/N$ cut imposed on the sample in this work, but are still within the uncertainties. The errors are estimated using 500 Monte Carlo realizations of the data, where we perturb the stellar parameters $T_{\rm eff}$, [Fe/H], and [$\alpha$/Fe] by the typical errors (200 K, 0.3 dex and 0.1 dex, respectively); see \S6 of \citetalias{cheng12a} for details.
 
\subsection{Kinematics of the High- and Low-$\alpha$ Populations}\label{kinematics}
In addition to different chemical properties, thin and thick disk stars in the solar neighborhood exhibit different kinematic properties. A comparison of the kinematics of high-$\alpha$ stars at large and small $R$ can help distinguish whether high-$\alpha$ stars at large $R$ are the outer disk tail of the inner disk population, the high-$\alpha$ tail of the outer disk population, or a different population altogether. In this section we examine the rotational velocities $V_{\phi}$ of high- and low-$\alpha$ stars as a function of $R$ and $|Z|$. For this analysis, we only consider \ngoodpm stars, which have good proper motions, as described below.

We calculate three-dimensional velocities in Cartesian coordinates ($U,V,W$), and polar coordinates ($V_R, V_{\phi}, V_Z$), using radial velocities along with proper motions obtained by combining the USNO-B and SDSS catalogs \citep{mun04}. We use the criteria of \citet{kil06} to obtain a sample with clean proper motions: \texttt{sigRa < 525 mas, sigDec < 525 mas, match = 1, nFit = 6, dist22 > 7\arcsec}, where \texttt{sigRa} and \texttt{sigDec} are the residual for the proper motion fit in Right Ascension and Declination, \texttt{match} is the number of objects within a 1\arcsec~radius, \texttt{nFit} is the number of plates the object was detected on, and \texttt{dist22} is the distance to the nearest neighbor with $g < 22$. The efficacy of these criteria have been explored by \citet{don11}. 

These selection criteria identify \ngoodpm stars in our sample with reliable proper motions; this subsample has the same distributions in distance, magnitude, color, [Fe/H], and [$\alpha$/Fe] as the larger sample of \ngoodafe stars, so we treat it as a representative sample and do not apply any additional weights. Statistical errors on the proper motions are roughly $3-3.5$ mas yr$^{-1}$, which corresponds to a tangential velocity error of $28-33$ km s$^{-1}$ at a distance of 2 kpc.

In this section we present histograms, corrected using our weighting scheme, which are needed to properly account for the different sampling along different lines of sight. For example, there are many more stars in the lines of sight toward smaller $R$ ($l < 90^{\circ}$) compared those at the anti-center ($l\sim180^{\circ}$), but the number of spectra are approximately equal in all directions. Accounting for this effect is necessary to reproduce the correct distributions, but doing so magnifies the Poisson noise.

\begin{figure*}[!t]
\epsscale{1.0}
\plotone{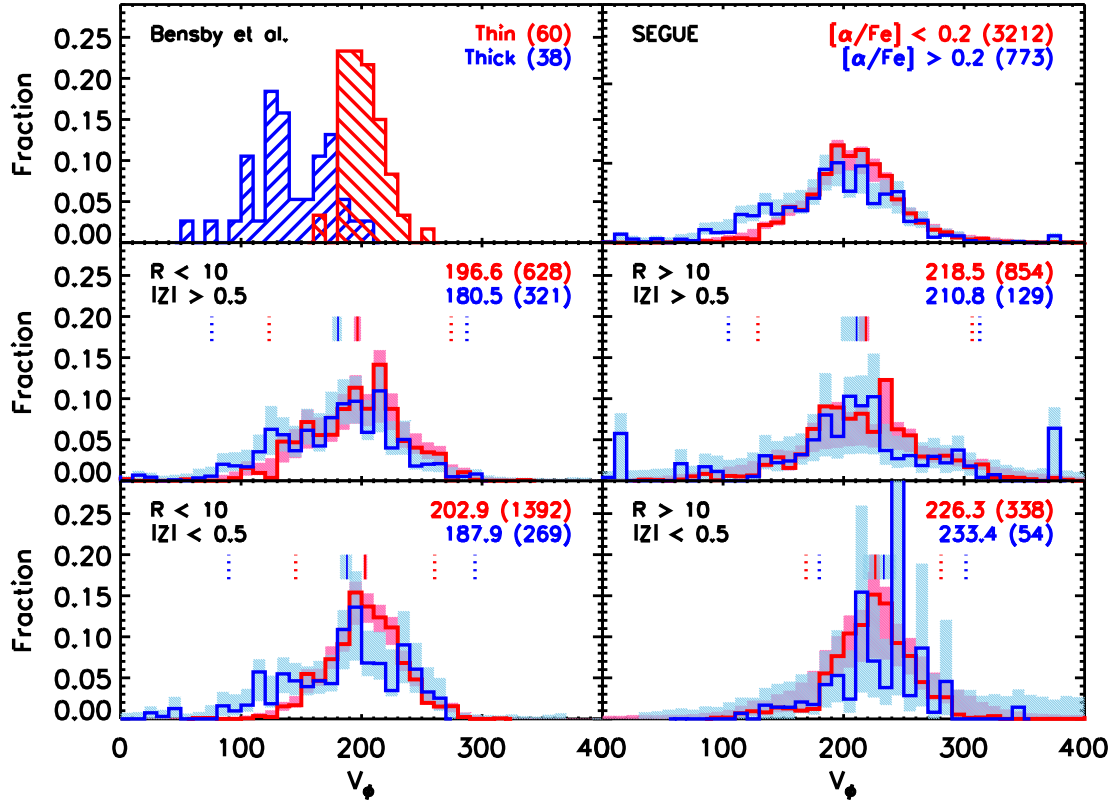}
\caption[Weighted rotational velocity $V_{\phi}$ distributions for high- and low-$\alpha$ stars.]{Weighted rotational velocity $V_{\phi}$ distributions for high- and low-$\alpha$ stars. Top left panel: the solar neighborhood sample of \citet{ben03,ben05}, with thin and thick disk stars (red and blue, respectively) assigned according to their kinematics. Top right panel: the SEGUE sample, with high- and low-$\alpha$ stars (blue and red, respectively) divided at [$\alpha$/Fe]$=+0.2$. The shaded regions indicate the errors estimated using 500 Monte Carlo realizations of our data. Bottom four panels: the SEGUE sample, divided into four bins of $R$ and $|Z|$. In the top right corner of each panel, the mean rotational velocity $\langle V_{\phi}\rangle$ of each population is indicated, with the raw number of stars in each population in parentheses. To calculate an outlier-resistant $\langle V_{\phi}\rangle$ (vertical solid lines), we exclude targets that are more than three median absolute deviations from the median value (vertical dotted lines). The errors on $\langle V_{\phi}\rangle$ are indicated by the surrounding shaded regions. At $R < 10$ kpc, the high-$\alpha$ stars lag in rotation behind the low-$\alpha$ stars by \lag. At $R > 10$ kpc, the difference in $\langle V_{\phi}\rangle$ between high- and low-$\alpha$ stars is $<8$ km s$^{-1}$, which is within the measurement errors. The different kinematic properties of high-$\alpha$ stars at large and small $R$ suggest that they may be different populations with different origins.}
\label{hist_vrot}
\end{figure*}

Figure~\ref{hist_vrot} shows $V_{\phi}$ histograms for the Bensby sample (top left panel), the total SEGUE sample (top right panel), and the SEGUE sample in the four $R, |Z|$ bins (bottom four panels), similar to Figure~\ref{hist_afe}. In each panel, as in Figure~\ref{hist_afe}, the sample is divided into high- and low-$\alpha$ stars (blue and red, respectively) at [$\alpha$/Fe]$=0.2$.  The shaded regions indicate errors estimated by generating 500 Monte Carlo realizations of our data, where we perturb the stellar parameters, radial velocities, and proper motions; the typical errors on the radial velocities and proper motions are 6.0 km s$^{-1}$ and 3 mas yr$^{-1}$, respectively. 

We calculate the mean rotational velocities $\langle V_{\phi}\rangle$ for each population (vertical solid lines) using an outlier-resistant algorithm, which trims values greater than three median absolute deviations from the median (vertical dotted lines in Figure~\ref{hist_vrot}). The numerical values of $\langle V_{\phi}\rangle$ are indicated in the top right corner of each panel, with the raw number of stars in parentheses. The errors on the means are indicated by the shaded regions surrounding the vertical solid lines. For the low-$\alpha$ samples, the error on the mean is $\sim2-3$ km s$^{-1}$, while for the high-$\alpha$ samples, which have fewer stars, it is $\sim4-10$ km s$^{-1}$. 

At $R < 10$ kpc, the kinematics of high-$\alpha$ stars are consistent with those seen for thick disk stars in the solar neighborhood. Figure~\ref{hist_vrot} shows that at all $|Z|$, they lag in rotation behind low-$\alpha$ stars by \lag, in rough agreement with measurements in the literature (e.g., \citealp{chi00,sou03,car10}), which have values $\sim20-50$ km s$^{-1}$, depending on how the thin and thick disk populations are separated. Like thick disk stars in the solar neighborhood, the high-$\alpha$ stars at $R < 10$ kpc also have wider distributions in $V_Z$ and $V_R$. This result implies that they belong to a kinematically ``hotter" population, which has larger random motions in the radial and vertical directions in addition to a larger lag in $V_{\phi}$. For reference, the mean, median, median absolute deviation, and skewness of the populations' distributions in $V_{\phi}$, $V_Z$, and $V_R$ for both high- and low-$\alpha$ stars are tabulated in Table~\ref{vel_stats}. The errors on $\langle V_{\phi}\rangle$ are indicated.

\begin{table*}
\centering
\caption{Properties of Velocity Distributions}
\scriptsize\begin{tabular}{lcccccc|cccccc}
\hline
& \multicolumn{6}{c}{$R < 10, |Z| > 0.5$} & \multicolumn{6}{|c}{$R > 10, |Z| > 0.5$}\\
\hline
& \multicolumn{2}{c}{$V_{\phi}$} & \multicolumn{2}{c}{$V_Z$} & \multicolumn{2}{c}{$V_R$} & \multicolumn{2}{|c}{$V_{\phi}$} & \multicolumn{2}{c}{$V_Z$} & \multicolumn{2}{c}{$V_R$}\\
\hline
& Low-$\alpha$ & High-$\alpha$ & Low-$\alpha$ & High-$\alpha$ & Low-$\alpha$ & High-$\alpha$ & Low-$\alpha$ & High-$\alpha$ & Low-$\alpha$ & High-$\alpha$ & Low-$\alpha$ & High-$\alpha$\\
\hline
Mean &  196.6$^{ +2.7}_{ -3.0}$ &  180.5$^{ +3.1}_{ -4.8}$ &   -2.4 &   -6.0 &  108.2 &   93.9 &  218.5$^{ +2.7}_{ -5.6}$ &  210.8$^{ +8.3}_{-13.1}$ &   18.8 &   21.8 &  -34.0 &  -37.2\\
Median$^{\rm b}$ &  198.8 &  181.5 &   -0.6 &   -8.1 &  106.0 &   93.3 &  217.8 &  208.7 &   16.6 &   20.7 &  -30.5 &  -40.4\\
MAD$^{\rm b,c}$ & 25.2 &   35.3 &   24.5 &   41.5 &   37.0 &   43.8 &   29.6 &   34.8 &   33.3 &   31.0 &   31.9 &   35.5\\
Skew &  -0.50$^{ +0.3}_{ -0.3}$ &  -1.77$^{ +0.3}_{ -0.2}$ &   0.08 &   0.34 &   0.20 &  -0.42 &   0.03$^{ +0.3}_{ -0.2}$ &  -0.49$^{ +0.8}_{ -0.4}$ &  -0.25 &  -0.07 &   1.77 &   0.23\\
\hline
& \multicolumn{6}{c}{$R < 10, |Z| < 0.5$} & \multicolumn{6}{|c}{$R > 10, |Z| < 0.5$}\\
\hline
& \multicolumn{2}{c}{$V_{\phi}$} & \multicolumn{2}{c}{$V_Z$} & \multicolumn{2}{c}{$V_R$} & \multicolumn{2}{|c}{$V_{\phi}$} & \multicolumn{2}{c}{$V_Z$} & \multicolumn{2}{c}{$V_R$}\\
\hline
& Low-$\alpha$ & High-$\alpha$ & Low-$\alpha$ & High-$\alpha$ & Low-$\alpha$ & High-$\alpha$ & Low-$\alpha$ & High-$\alpha$ & Low-$\alpha$ & High-$\alpha$ & Low-$\alpha$ & High-$\alpha$\\
\hline
Mean$^{\rm a,b}$ & 202.9$^{+1.7}_{-1.3}$ &  187.9$^{+4.3}_{-5.8}$ &    1.1 &   -3.9 &   50.0 &   46.1 &  226.3$^{+3.8}_{-5.0}$ &  233.4$^{+10.2}_{-17.3}$ &   16.1 &   18.2 &  -41.1 &  -25.6\\
Median$^{\rm b}$ &  203.0 &  192.0 &    2.3 &   -6.5 &   52.9 &   46.3 &  224.9 &  240.7 &   15.1 &   11.7 &  -38.6 &  -10.4\\
MAD$^{\rm b,c}$ &   19.2 &   34.1 &   17.0 &   24.9 &   35.8 &   45.2 &   18.7 &   20.3 &   20.1 &   15.9 &   22.6 &   23.2\\
Skew &  -0.20$^{ +0.2}_{ -0.3}$ &  -1.27$^{ +0.4}_{ -0.3}$ &  -0.13 &  -0.04 &  -0.18 &  -0.80 &  -0.33$^{ +0.6}_{ -0.3}$ &  -0.70$^{ +1.2}_{ -0.5}$ &   0.06 &  -0.07 &  -0.28 &  -0.98\\
\hline
\label{vel_stats}
\end{tabular}
\\$^{\rm a}$ Outliers clipped at three median absolute deviations
\\$^{\rm b}$ Mean, median, and MAD reported in km s$^{-1}$.
\\$^{\rm c}$ Median Absolute Deviation
\end{table*}

In addition, there is a large fraction of high-$\alpha$ stars with low $V_{\phi}$ ($< 150$ km s$^{-1}$); this tail is not present in the low-$\alpha$ population. One way to quantify this feature is to examine the skewness of the $V_{\phi}$ distributions of both populations. We find that for the high-$\alpha$ population, the skewness of the $V_{\phi}$ distribution is larger than for the low-$\alpha$ population; at $|Z| < 0.5$ kpc ($|Z| > 0.5$ kpc) the $V_{\phi}$ distribution of the high-$\alpha$ population has a skewness of $-1.77^{ +0.3}_{ -0.2}$ ($-1.27^{ +0.4}_{ -0.3}$), while the $V_{\phi}$ distribution of the low-$\alpha$ population has a skewness of $-0.50^{ +0.3}_{ -0.3}$ ($-0.20^{ +0.2}_{ -0.3}$).

The skewed shape of the $V_{\phi}$ distribution of both high- and low-$\alpha$ stars at $R < 10$ kpc is consistent with a population that is falling in density with increasing $R$. If stars that have slow (fast) rotational velocities are on orbits with guiding centers within (beyond) the solar circle, then they are interlopers from the inner (outer) disk (e.g., \citealp{bin98}); the skewness of the distribution results from the higher stellar densities in the inner disk compared to the outer disk. Thus the larger skewness of the high-$\alpha$ population in $V_{\phi}$ is consistent with there being a steeper density gradient for high-$\alpha$ stars compared to that of low-$\alpha$ stars. The steeper density gradient of the high-$\alpha$ stars is consistent with the picture of the high-$\alpha$ population having a short scale length compared to the low-$\alpha$ population.

At $R > 10$ kpc, however, high- and low-$\alpha$ stars have the same mean rotational velocities, within the errors, and the fraction of high-$\alpha$ stars with low $V_{\phi}$ is comparable to that of the low-$\alpha$ stars. In addition, the widths of the $V_Z$ and $V_R$ distributions of high- and low-$\alpha$ stars are similar. While at $R < 10$ kpc, the different $V_{\phi}$ distributions are indicative of two populations with different structural parameters, at $R > 10$ kpc, we are unable to distinguish between the kinematics of high- and low-$\alpha$ stars. If high- and low-$\alpha$ stars at $R > 10$ kpc are part of the same population, then the observations imply that high-$\alpha$ stars at large and small $R$ may have different origins.

\section{Scale Length of the High-$\alpha$ Population}\label{measurescalelengths}
The lack of high-$\alpha$ stars at large $R$ suggests that the high-$\alpha$ population, which is typically associated with the thick disk, has a short radial extent. The similarity between the $V_{\phi}$ distributions of high- and low-$\alpha$ stars at large $R$ also supports this idea. In this section, we estimate the scale length of the $\alpha$-enhanced thick disk by quantifying the fraction of low- and high-$\alpha$ stars as a function of $R$ and $|Z|$. We then compare the data to expected values based on different combinations of thin and thick disk scale lengths, $L_{\rm thin}$ and $L_{\rm thick}$. In addition, we require that the predicted total stellar density as a function of $R$ and $|Z|$ is consistent with the total stellar density of the best-fit model obtained by \citet{jur08}. While it is possible that the data may be better fit by different analytical functions, we restrict our analysis to a radial exponential profile for the disk, for which we also have total stellar density measurements.

In using the fractions of high- and low-$\alpha$ stars to estimate the thin and thick disk scale lengths, we implicitly assume that high-$\alpha$ stars belong to the thick disk and low-$\alpha$ stars belong to the thin disk, as is observed in the solar neighborhood. We include the second constraint that the total stellar density match that of \citet{jur08} because our data provide a poor constraint on the local normalization of the disk models we fit. In our fits, we fix the scale heights and the total normalization to the \citet{jur08} values. We exclude the lowest $|Z|$ slice, where our sample does not cover a large range in $R$. In this section, we present the results of our analysis; details about the our procedure and the uncertainties in our estimates are provided in the Appendix.

\begin{figure*}[!t]
\epsscale{1.0}
\plotone{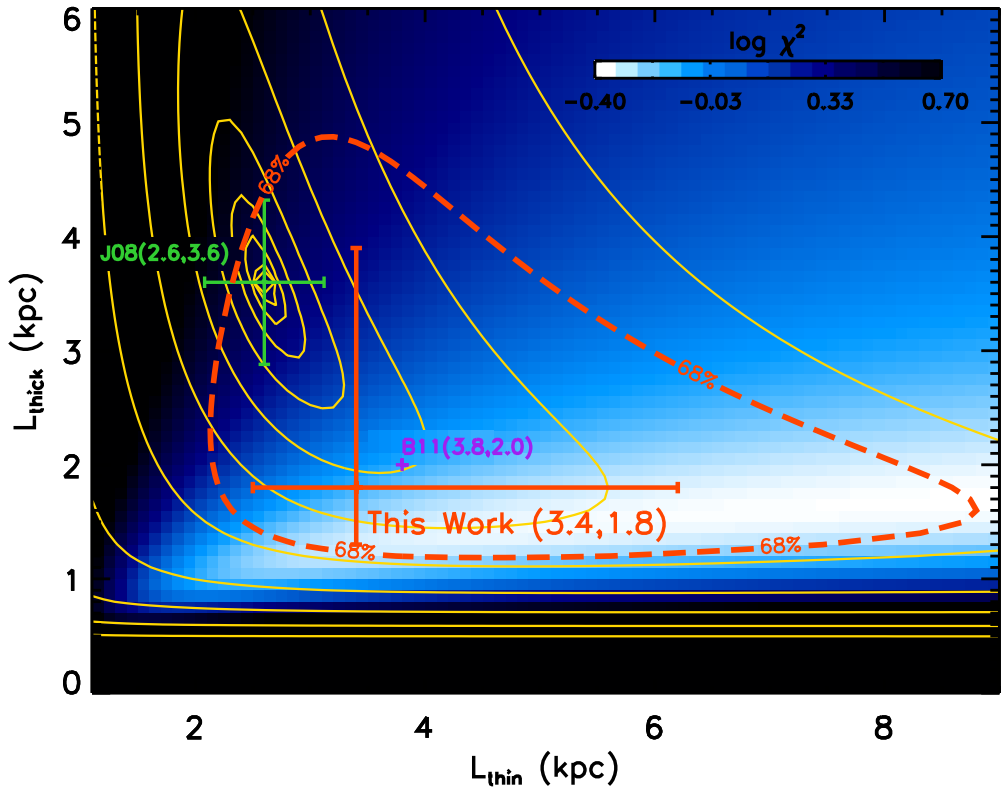}
\caption[Best-fit thin and thick disk scale lengths.]{Best-fit thin and thick disk scale lengths. The blue shading shows the reduced $\chi^2$ values using the fractions of high- and low-$\alpha$ stars as a function of $R$ and $|Z|$ as a constraint. The thin yellow contours show the constraint of the total stellar density as determined by \citet{jur08}. Our best estimate of the thin and thick disk scale lengths using both constraints are $L_{\rm thin}=\lthin\lthinerr$ kpc and $L_{\rm thick}=\lthick\lthickerr$ kpc, marked by the large orange cross. The 68\% contour for the combined constraint is shown as the thick dashed orange line. The published values of \citet{jur08} and \citet{ben11} are indicated in green and purple, respectively.}
\label{scalelengths}
\end{figure*}

The blue shading in Figure~\ref{scalelengths} represents the reduced $\chi^2$ values we obtain by comparing the observed and expected values of the high- and low-$\alpha$ fractions, as a function of $R$ and $|Z|$, for different combinations of $L_{\rm thin}$ and $L_{\rm thick}$. The thin yellow contours show the discrepancy between the total stellar density predicted by each combination of scale lengths and the total density from the best-fit scale lengths reported by \citet[see also their Figure 21]{jur08}. Combining both the constraints of total stellar density from \citet{jur08} and the high- and low-$\alpha$ fractions from our sample, we find the best combination of thin and thick disk scale lengths to be $L_{\rm thin}=\lthin\lthinerr$ kpc, $L_{\rm thick}=\lthick\lthickerr$ kpc, marked by the large orange cross in Figure~\ref{scalelengths}. The best fit values and error bars are obtained by marginalizing over each axis and determining the 68\% confidence interval for each scale length. The 68\% contour in two dimensions for the combined constraint is shown as the thick dashed orange line. More details about the procedure and error analysis are given in the Appendix. The blue-shaded map shows that for any given thin disk scale length, the preferred thick disk scale length, as constrained by the high- and low-$\alpha$ fractions (i.e., the white regions), is always shorter thin disk scale length. 

We note that our measurement of the thin and thick disk scale lengths is not well constrained, as we do not have data at large radius near the midplane. Our preferred value for the thin disk scale length is slightly larger, but consistent with values based on near-infrared photometry from the Spacelab Infrared Telescope \citep[$\sim3.0$ kpc]{ken91} and the Cosmic Background Explorer \citep[$\sim2.6$ kpc]{fre98}. Other SDSS/SEGUE studies, which use detailed modeling to estimate disk structural parameters, are also consistent with these values for the thin disk scale length \citep{jur08,bov11}. While our sample's spatial coverage is not ideal for constraining the thick disk scale length, it is an improvement on earlier studies. Furthermore, our best-fit value does not change significantly when we vary the assumed \citet{jur08} scale lengths or remove possible halo contaminants. See the Appendix for more details.

\begin{figure*}[!t]
\epsscale{1.0}
\plotone{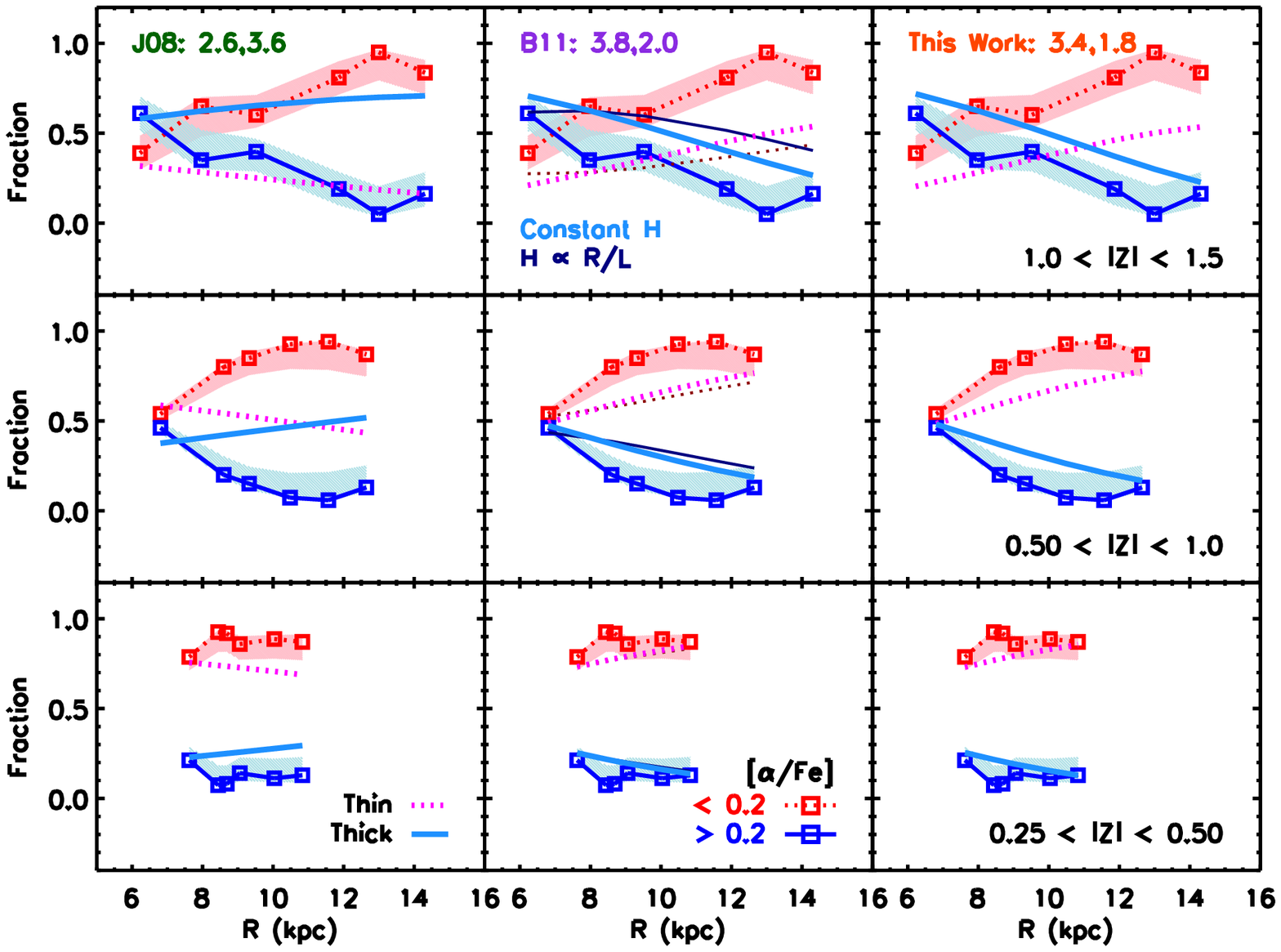}
\caption[Weighted fractions of high- and low-$\alpha$ stars vs. Galactocentric radius $R$ in three $|Z|$ slices.]{Weighted fractions of high- and low-$\alpha$ stars (blue and red squares, respectively) vs. Galactocentric radius $R$ in three $|Z|$ slices. The shaded regions indicate the errors on the measured fractions. Pink dotted and blue solid lines show the expected contributions of the thin and thick disks, respectively, for different combinations of scale lengths: (1) $L_{\rm thin}=2.6$, $L_{\rm thick}=3.6$ kpc in the left column \citep{jur08}, (2) $L_{\rm thin}=3.8$, $L_{\rm thick}=2.0$ kpc in the middle column \citep{ben11}, and (3) $L_{\rm thin}=\lthin$, $L_{\rm thick}=\lthick$ kpc in the right column (this work). In the middle column, we also show the expected fractions if the scale height varies linearly with $R_G/L$, as discussed by \citet[thinner lines]{ben11}. Our data support a shorter scale length for the high-$\alpha$ population, in agreement with the conclusion of \citet{ben11}.}
\label{radial_fraction}
\end{figure*}

Figure~\ref{radial_fraction} shows the weighted fractions of high- and low-$\alpha$ stars (blue and red squares, respectively), as a function of $R$ in three $|Z|$ slices. The shaded regions indicate the errors on the measured fractions. Each column shows the same data compared to the expected fractions of thin (pink dotted lines) and thick (blue solid lines) disk stars for three different combinations of thin and thick disk scale lengths: (1) $L_{\rm thin}=2.6$, $L_{\rm thick}=3.6$ kpc in the left column \citep{jur08}, (2) $L_{\rm thin}=3.8$, $L_{\rm thick}=2.0$ kpc in the middle column \citep{ben11}, and (3) $L_{\rm thin}=\lthin$, $L_{\rm thick}=\lthick$ kpc in the right column (this work). For the \citet{ben11} scale lengths, we show the predictions for both a constant thick disk scale height (thin, dark blue line) and one that varies as $R_G/L$ (thick, light blue line), as described in their paper. 

As seen in Figures~\ref{radial_zbin} and \ref{hist_afe}, Figure~\ref{radial_fraction} shows that the fraction of high-$\alpha$ stars decreases at large $R$, in every $|Z|$ slice. If we assume that the thick disk is populated only by high-$\alpha$ stars, then the model of \citet{jur08} vastly overpredicts the fraction of high-$\alpha$ stars at $|Z| > 0.5$ kpc; in our bin $R > 10$ kpc, $|Z| < 0.5$ kpc (Figure~\ref{hist_afe}), we would expect $\sim50\%$ of our sample ($\sim700$ stars) to be enhanced in [$\alpha$/Fe] instead of the $\sim10\%$ ($\sim200$ stars) that we observe. The data agree much better with a shorter thick disk scale length, consistent with the results of \citet[middle column]{ben11}. Using [$\alpha$/Fe] as a proxy for membership in the thin and thick disks in our larger sample, we estimate that the thick disk has a shorter scale length than the thin disk (right column).

\section{Discussion}\label{discussion}
\subsection{The Thick Disk Scale Length}
The results presented above show that the fraction of high-$\alpha$ stars drops off at large $R$ and that the high-$\alpha$ populations at small and large $R$ have different kinematic properties. Both of these results are consistent with the properties of the K-giant sample of \citet{ben11}, who found that the lack of high-$\alpha$ stars was consistent with the thick disk having a shorter scale length than the thin disk ($L_{\rm thin}=3.8$, $L_{\rm thick}=2.0$ kpc). Using our data, we estimate the scale lengths to be $L_{\rm thin}=\lthin\lthinerr$ and $L_{\rm thick}=\lthick\lthickerr$ kpc. While the scale lengths are not well constrained with our data, the thick disk scale length is consistently found to be shorter than 2 kpc (see the Appendix).

{\color{black} Independent analyses with other SDSS/SEGUE samples have also found similar scale lengths for stars with thick disk chemistry. \citet{car10}, using the velocity ellipsoid of stars in a particular metallicity range and location in the disk ($-0.8 <$ [Fe/H] $< -0.6, 1 < |Z| < 2$ kpc), estimated the thick disk scale length to be 2.2 kpc. In recent work, \citet{bov11} measured the scale length of their ``$\alpha$-old" population ($-1.5 <$ [Fe/H] $< -0.25, 0.25 < $ [$\alpha$/Fe] $<0.50$) to be 1.96 kpc. While the methods differ, these studies reach the same conclusion: the population of stars associated with a thick disk component in the solar neighborhood has a short radial scale length.

It is worth emphasizing, however, that the above results apply for stars with particular properties. The results presented in this work reflect the \textit{fractions} of high- and low-$\alpha$ stars,} and our scale length estimate reflects the radial extent of the high-$\alpha$ population, which we associate with the thick disk based on studies of the solar neighborhood. Previous studies of external galaxies \citep{dal02,yoa06} and the Milky Way \citep{jur08,dej10} have relied on surface brightness profiles and star counts, respectively, which follow the total stellar density, with no information about the stellar populations.

We now introduce terminology to distinguish between these two methods of identifying the two components of the disk. First, we will refer to the \textit{structural} thin and thick disks to describe the components that are identified using the total stellar densities, either through star counts or surface brightness profiles (e.g., \citealt{gil83,dal02}). Second, we will refer to the \textit{chemical} thin and thick disks to describe the components that are identified using the chemical and/or kinematic properties of stars (e.g., \citealt{ben03,ben05,lee11b}). In our work, we have found that the \textit{chemical} thick disk has a shorter scale length than the \textit{chemical} thin disk.

Whether \textit{star counts} in the SEGUE imaging along our lines of sight support a short scale length for the \textit{structural} thick disk remains an open question. Previous estimates of the scale length of the structural thick disk in the Milky Way have typically relied on star counts in higher latitude data and therefore do not have significant leverage in the radial direction (e.g., \citealt{jur08,cha11}). Our lines of sight reach larger $R$ at small $|Z|$ and may provide additional constraints on the scale lengths of the structural thin and thick disks. Comparing star counts in our low latitude lines of sight to different combinations of structural parameters, as well as exploring different radial profiles, will be the focus of future work.

If the structural thick disk is found to have a short scale length, in agreement with our result for the chemical thick disk, then we can compare these results to the scale lengths of structural thick disks seen in external galaxies. Observationally, structural thin and thick disk scale lengths have been found to be uncorrelated in external galaxies \citep{dal02}. For galaxies with circular velocities greater than $\sim$100 km s$^{-1}$, structural thin disks with larger scale lengths than structural thick disks have been reported (see Table 5 of \citealp{yoa06}). Having a structural thick disk with a short scale length, then, would not make the Milky Way an unusual galaxy.

\subsection{Possible Relation to the Hercules Thick Disk Cloud}
In Table~\ref{los_fraction}, we list the weighted fraction of high-$\alpha$ stars along each of the 11 lines of sight in our sample. Most of the 11 lines of sight have fewer than $15\%$ of their stars with [$\alpha$/Fe]~$>+0.2$. Three lines of sight at $R < 10$ kpc, however, have $\sim20-50\%$ of their stars with [$\alpha$/Fe]~$>+0.2$. These lines of sight are directed toward the Hercules Thick Disk Cloud, a stellar overdensity in the disk, which has been studied in detail by \citet{lar10,lar11} and \citet{hum11}. 

\begin{table*}
\centering
\caption{Fraction of High-$\alpha$ Stars per Line of Sight}
\scriptsize\begin{tabular}{llrrrrrr}
\hline
\multicolumn{2}{c}{Plug-Plates} & \multicolumn{1}{c}{$l$ ($^{\circ}$)} & \multicolumn{1}{c}{$b$ ($^{\circ}$)} & N$_{\rm stars}$ & N$_{\rm high-\alpha}$ & f$_{\rm unweighted}$ & f$_{\rm weighted}$ \\
\hline
2534 & 2542 &   50.0 &   14.0 & 396 & 210 &  0.530 &  0.486\\
2536 & 2544 &   70.0 &   14.0 & 401 & 155 &  0.387 &  0.305\\
2554 & 2564 &   94.0 &   14.0 & 600 & 165 &  0.275 &  0.212\\
2555 & 2565 &   94.0 &    8.0 & 386 & 40 &  0.104 &  0.103\\
2556 & 2566 &   94.0 &   -8.0 & 526 & 68 &  0.129 &  0.092\\
2538 & 2546 &  110.0 &   16.0 & 553 & 92 &  0.166 &  0.143\\
2537 & 2545 &  110.0 &   10.5 & 467 & 56 &  0.120 &  0.093\\
2681 & 2699 &  178.0 &  -15.0 & 495 & 82 &  0.166 &  0.132\\
2668 & 2672 &  187.0 &  -12.0 & 670 & 81 &  0.121 &  0.094\\
2678 & 2696 &  187.0 &    8.0 & 580 & 67 &  0.116 &  0.080\\
2712 & 2727 &  203.0 &    8.0 & 546 & 93 &  0.170 &  0.142\\
\hline
\label{los_fraction}
\end{tabular}
\end{table*}

\citet{lar10} detect this overdensity as an excess in star counts in the first quadrant (Q1, $0^{\circ} < l < 90^{\circ}$), compared to the fourth quadrant (Q4, $270^{\circ} < l < 360^{\circ}$), in particular, at Galactic coordinates $20^{\circ} < l < 55^{\circ}, 20^{\circ} < b < 45^{\circ}$. Stars associated with this overdensity in Q1 lag in rotation by 30 km s$^{-1}$ compared to stars in Q4, but have metallicities similar to stars in analogous fields in Q4 \citep{par04,hum11}. Their preferred scenario for the existence of the overdensity is that dynamical interactions with the bar cause stars to pile up in a ``gravitational wake" (e.g., \citealt{her92}). This feature may be related to the Hercules-Aquila Cloud seen in the SDSS \citep{bel07,jur08}.

The rotation rates, $\omega$, in our three high-$\alpha$ lines of sight ($22-31$ km s$^{-1}$ kpc$^{-1}$) are slightly larger than the Q1 fields of \citet[$15-26$ km s$^{-1}$ kpc$^{-1}$]{hum11}, which is not unexpected, since our lines of sight are at lower Galactic latitude. A direct comparison is not possible because our samples do not overlap spatially. Additionally, our sample does not cover a sufficiently large part of the Galaxy to fully test the presence of an asymmetry: all of our inner disk stars are in Q1, and we have no stars in Q4 with which to make a comparison. We do not currently have the necessary data to confirm or exclude the possibility that these three lines of sight are associated with this overdensity.

\subsection{Implications for Thick Disk Formation}
The short scale length of the chemical thick disk can be used to constrain various scenarios for thick disk formation, such as the four described in \S\ref{intro}. In the following discussion, we will use the generic term ``thick disk" to refer to both the structural and chemical thick disks. We assume these to be the same, as is done in all of the models discussed.

The lack of high-$\alpha$ stars at $R > 10, |Z| > 0.5$ kpc puts an upper limit on the strength of migration due to transient spiral structure (scenario 4); the $N$-body simulation of \citet{loe11}, for example, predict that this mechanism can transport many high-$\alpha$ stars from the inner disk to large $R$ and $|Z|$. In this simulation, high-$\alpha$ stars are present at all $R$ because they are old and have had more time to migrate to large $R$. The lack of high-$\alpha$ stars that we have observed at large $R$, then, implies that this mechanism cannot be very efficient.

If the stars we observe at large $R$ and $|Z|$ reached their current positions through radial migration, the limited extent of the high-$\alpha$ population could be evidence that the mechanism must have some radial dependence on the strength of migration. One such mechanism relies on the presence of a bar and a steady state spiral pattern, which leads to more efficient mixing at certain radii (e.g., \citealp{min10,bru11}). \citet{bru11}, for example, find that stars that are close to the corotation radius of the bar are more likely to migrate. If the high-$\alpha$ stars in our sample were born in the bulge and have since migrated to where we observe them at $R < 10$ kpc, this would explain the similar abundance patterns that have been reported for thick disk and bulge stars \citep{mel08,ben09,alv10,ben10a,ryd10,gon11}. One way to test this scenario is to examine whether the kinematic properties of stars in these simulations are different for those mixed in the inner and outer disks.

If the thick disk does indeed have a shorter scale length than the thin disk, as is suggested by both our data and the data of \citet{ben11}, and radial migration is not the dominant mechanism, our results may have implications on the formation and merger history of the Milky Way disk. A range of thick disk scale lengths can result from different merger histories. \citet{bro04,bro07} showed that chaotic gas accretion at early times (scenario 3) results in a thick disk with a shorter scale length than the thin disk, while an early gas-rich 2:1 merger of two disks results in a thick disk with a longer scale length. The variation in merger histories would provide a possible explanation for the range of structural thin and thick disk scale lengths observed in nearby galaxies \citep{dal02,yoa06}. 

Scenarios involving minor mergers can also be constrained. If thick disk stars originated from an initially thin disk (scenario 1) then any heating event must have occurred in a primordial disk with a short scale length. The predominantly low-$\alpha$ stars at large $R$ and $|Z|$, then, should come from a more chemically evolved disk. Radial mixing induced by late minor mergers has been shown to be very efficient for stars in the outer disk \citep{bir12} and could explain the presence of low-$\alpha$ stars at large $R$ and $|Z|$. However, predictions can be dependent on the particular models being examined and their initial conditions (e.g., \citealp{di-11}). If thick disk stars originated from an accreted satellite (scenario 2), then stars contributed by a single satellite should be located in a torus-like structure, and we should see the same abundance trends in Q1 and Q4. Our current sample is insufficient in spatial coverage to explore this possibility.

Finally, the radial gradients in [Fe/H] for high- and low-$\alpha$ stars (see Figures~\ref{alphas_rfeh} and~\ref{alphas_slopes}) also provide constraints on thick disk formation. Radial migration due to transient spiral structure (scenario 4) could explain the flattening trend in the low-$\alpha$ stars, but the mechanism is too efficient in current simulations, resulting in too many high-$\alpha$ stars in the outer disk, especially at large $|Z|$. The observed distributions of [Fe/H] and [$\alpha$/Fe] could be explained by the following: First, early gas-rich mergers (scenario 3) created a chemically homogeneous, high-$\alpha$ population in a thick disk with a short scale length. Subsequently, a low-$\alpha$ thin disk forms and is heated by minor merger activity at later times (scenario 1), mixing stars at large $R$ (e.g., \citealt{bir12}) and flattening the radial metallicity gradient at large $|Z|$.

\section{Summary}
We have demonstrated, using a sample of \ngoodafe main sequence turnoff stars from the SEGUE survey, that the high-$\alpha$ population, which is associated with the thick disk in the solar neighborhood, has a short scale length ($L_{\rm thick}\sim\lthick$ kpc) and a flat metallicity gradient at all $|Z|$. The abundance trends at $R < 10$ kpc show a dichotomy between high- and low-$\alpha$ stars similar to that seen between thick and thin disk stars observed in the solar neighborhood \citep{ben03,ben05}. The fraction of high-$\alpha$ stars increases with $|Z|$, and these high-$\alpha$ stars lag in rotation compared to low-$\alpha$ stars (by \lag), similar to the difference in kinematics seen for thin and thick disk stars in the solar neighborhood \citep{chi00,sou03}. 

At $R > 10$ kpc, the fraction of high-$\alpha$ stars is lower than at small $R$ and does not increase with $|Z|$. High-$\alpha$ stars at large $R$ also do not lag in rotation compared to low-$\alpha$ stars, with both populations having similar mean rotational velocities. These results suggest that the high-$\alpha$ stars in the outer disks may simply be the tail of the [$\alpha$/Fe] distribution; the stars far from the plane in the outer disk ($R > 10, |Z| > 0.5$ kpc) likely have different origins than those far from the plane in the inner disk ($R < 10, |Z| > 0.5$ kpc). 

The fractions of high- and low-$\alpha$ stars are consistent with the expected values for a thick disk with a short scale length as suggested by \citet{ben11}. Using the fractions of high- and low-$\alpha$ stars as a function of $R$ and $|Z|$, we estimate the thick disk scale length to be $L_{\rm thick}\sim\lthick$ kpc. In addition, it is possible that the lines of sight in our sample with large fractions of high-$\alpha$ stars are related to the Hercules Thick Disk Cloud, a stellar overdensity studied by \citet{hum11}. A sample of stars with better spatial coverage, particularly in Q4, is required to fully explore this possibility.

We find that the presence of a thick disk with a short scale length is consistent with the scenario of \citet{bro04,bro05}, in which the thick disk formed during a turbulent disk phase at early times when gas accretion rates were high. In the outer disk, stars may have been moved to large $R$ and $|Z|$ through radial mixing due to late minor mergers (e.g., \citealt{bir12}). The lack of high-$\alpha$ stars can be used to constrain the strength of radial migration of stars from the inner disk induced by transient spiral structure (e.g., \citealp{ros08a,ros08b,sch09a,sch09b,loe11}). If stars in the outer disk arrived at their current locations through radial migration, some radially-dependent mechanisms may be responsible (e.g., \citealt{min10,bru11}).

\acknowledgments
We would like to thank the anonymous referee for useful suggestions and thoughtful comments. J.Y.C. would also like to thank S. Loebman, R. Sch\"{o}nrich, A. Robin, J. Bovy, and A. S. Lee for useful conversations and comments. C.M.R. gratefully acknowledges funding from the David and Lucile Packard Foundation, and thanks the Max-Planck-Institute f\"{u}r Astronomie (MPIA), Heidelberg for hospitality. Y.S.L. and  T.C.B. acknowledge partial funding of this work from grants PHY 02-16783 and PHY 08-22648: Physics Frontier Center/Joint Institute for Nuclear Astrophysics (JINA), awarded by the U.S. National Science Foundation.

Funding for SDSS-III has been provided by the Alfred P. Sloan Foundation, the Participating Institutions, the National Science Foundation, and the U.S. Department of Energy Office of Science. The SDSS-III web site is http://www.sdss3.org/. SDSS-III is managed by the Astrophysical Research Consortium for the Participating Institutions of the SDSS-III Collaboration including the University of Arizona, the Brazilian Participation Group, Brookhaven National Laboratory, University of Cambridge, Carnegie Mellon University, University of Florida, the French Participation Group, the German Participation Group, Harvard University, the Instituto de Astrofisica de Canarias, the Michigan State/Notre Dame/JINA Participation Group, Johns Hopkins University, Lawrence Berkeley National Laboratory, Max Planck Institute for Astrophysics, Max Planck Institute for Extraterrestrial Physics, New Mexico State University, New York University, Ohio State University, Pennsylvania State University, University of Portsmouth, Princeton University, the Spanish Participation Group, University of Tokyo, University of Utah, Vanderbilt University, University of Virginia, University of Washington, and Yale University. 

Facilities: \facility{Sloan}.

\appendix
\section{Appendix: Scale Length Estimates}\label{appendix}
\section{Procedure}
To determine the scale lengths of the thin and thick disks, we follow the prescription of \citet{jur08} and model the Galaxy as two double exponential disks plus a two-axial power-law ellipsoid halo (their Equations 21-24):
\begin{eqnarray}
\rho(R,Z) = \rho_{\rm D}(R,Z;L_{\rm thin},H_{\rm thin}) \nonumber\\
+ f\rho_{\rm D}(R,Z;L_{\rm thick},H_{\rm thick}) + \rho_{\rm H}(R,Z)
\end{eqnarray}
where
\begin{eqnarray}
\rho_{\rm D}(R,Z;L,H) = \rho_{\rm D}(R_{\odot},0)\times{\rm exp}\left(\frac{R_{\odot}}{L}\right) \nonumber \\
\times {\rm exp}\left(-\frac{R}{L}-\frac{Z+Z_{\odot}}{H}\right)
\end{eqnarray}
and
\begin{eqnarray}
\rho_{\rm H}(R,Z) = \rho_{\rm D}(R_{\odot},0)f_{\rm H}\left[\frac{R_{\odot}}{\sqrt{R^2+(Z/q_{\rm H})^2}}\right]^{n_{\rm H}}
\end{eqnarray}

The definitions of the parameters and the values used (the bias-corrected parameters in their Table 10) are listed in Table~\ref{juricfit}. We vary the thin and thick disk scale lengths and fix the remaining parameters (including the scale heights and the total normalization) to the \citet{jur08} values, which are constrained using their photometric sample. Our spectroscopic sample is not well suited for determining the total stellar density because of the smaller sample size and the pencil-beam nature of the observations. 

For each combination of scale lengths, we calculate a reduced $\chi^2$ statistic to indicate how well the predicted fractions of high- and low-$\alpha$ stars as a function of $R$ and $|Z|$ reproduce what we see in our SEGUE sample; these values are indicated by the blue-shaded map in Figure~\ref{scalelengths}. In addition, we calculate how well the sum of the two exponential disks matches the total density measured by \citet{jur08}; these values are indicated by the thin yellow contours in Figure~\ref{scalelengths}. This second constraint is not strictly a $\chi^2$ statistic, as we are comparing two smooth models. We normalize the second constraint such that it has the same 10th and 90th percentile levels as the $\chi^2$ values from the first constraint.

We determine our best-fit scale lengths by calculating a probability for each combination of scale lengths, where the probability is proportional to $e^{-\chi^2/2}$. All probabilities are normalized such that the total probability in the parameter space $1 < L_{\rm thin} < 10$ kpc, $0 < L_{\rm thick} < 8$ kpc is equal to one. The thick dashed orange contour in Figure~\ref{scalelengths} shows the contour that encompasses 68\% of the volume under the surface defined by the probabilities. The best fit value of each scale length is obtained by marginalizing the probabilities over each axis and determining the peak in one dimension. The error bars indicate the 68\% confidence interval in one dimension. This exercise yields our final results for the scale lengths: $L_{\rm thin}=\lthin\lthinerr$ kpc, $L_{\rm thick}=\lthick\lthickerr$ kpc.

\begin{table*}
\centering
\caption{Structural Parameters Measured by \citet{jur08}}
\scriptsize\begin{tabular}{lcrl}
\hline
Parameter & Bias-Corrected & Error & Definition\\
\hline
$Z_{\odot}$ & 25 pc & 20\% & Solar offset from the Galactic plane\\
$L_{\rm thin}$ & 2600 pc & 20\% & Thin disk scale length\\
$H_{\rm thin}$ & 300 pc & 20\% & Thin disk scale height\\
$f$ & 0.12 & 10\% & Thick disk normalization relative to thin disk at $R=R_{\odot}, Z = 0$\\
$L_{\rm thick}$ & 3600 pc & 20\% & Thick disk scale length\\
$H_{\rm thick}$ & 900 pc & 20\% & Thick disk scale height\\
$f_{\rm H}$ & 0.0051 & 25\% & Halo normalization relative to thin disk at $R=R_{\odot}, Z = 0$\\
$q_{\rm H}$ & 0.64 & $\lesssim0.1$ & Halo ellipticity\\
$n_{\rm H}$ & 2.77 & $\lesssim0.2$ & Halo power law\\
\hline
\label{juricfit}
\end{tabular}
\end{table*}

\section{Additional Sources of Error}
In this section, we estimate the errors in our scale length estimates due to (1) random errors in the stellar parameters, (2) systematic errors in the \citet{jur08} scale lengths, and (3) contamination by halo stars. First, we estimate the random errors on the scale lengths using the same method as in \citetalias{cheng12a}, where we generate 500 Monte Carlo realizations of our data (see \S6.3 of \citetalias{cheng12a} for more details). In each realization, we perturb the stellar parameters $T_{\rm eff}$, [Fe/H], and [$\alpha$/Fe] by the typical errors (200 K, 0.3 dex and 0.1 dex, respectively). We find that errors in the stellar parameters only change the scale lengths by 0.1 kpc.

Secondly, to estimate the errors in the assumed total density (i.e., the thin yellow contours in Figure~\ref{scalelengths}), we repeat the calculation, varying the \citet{jur08} scale lengths and scale heights by their reported errors ($20\%$). The purpose of this exercise is to simulate the effect of systematic errors between the distances of \citet{jur08} and the present work, which will cause the structural parameters to increase or decrease together. When we increase the \citet{jur08} values by $20\%$, we obtain $L_{\rm thin}=8.1^{+1.3}_{-2.7}$ kpc, $L_{\rm thick}= 1.8^{+3.7}_{-0.6}$ kpc; for a $20\%$ decrease, we obtain $L_{\rm thin}=2.5^{+1.8}_{-0.6}$ kpc, $L_{\rm thick}=1.7^{+1.4}_{-0.5}$ kpc. 

Lastly, to test the robustness of our results, we repeat the calculation after removing stars that may belong to the halo. Halo stars also have a short scale length and are enhanced in [$\alpha$/Fe]. We adopt the three criteria to identify probable halo stars: (1) a metallicity cut that removes all stars with [Fe/H] $<-0.7$, (2) a kinematic cut, which removes all stars with $V_{\phi} < 150$ km s$^{-1}$, and (3) a kinematic cut that removes all stars with $V_{\rm Gal} < 100$ km s$^{-1}$ to remove stars with the largest velocity offset relative to the projection of the local standard of rest, where $V_{\rm Gal} = V_R + 220\cdot{\rm cos}b\cdot{\rm sin}l$ and $V_R$ is the line-of-sight velocity measured from the SEGUE spectra. We only remove stars with $V_{\rm Gal} < 100$ km s$^{-1}$ along lines of sight with $50 < l < 130^{\circ}$. We do not include the lines of sight directed toward the Galactic anticenter because the local standard of rest is tangent to those directions, and the projection does not give a meaningful velocity. For all three criteria we obtain the same scale lengths, which suggests that halo contamination does not affect our scale length measurements.

The above analysis shows that the best-fit thick disk scale length is not significantly affected by errors in the stellar parameters, our assumptions of the total stellar density, and possible contamination from halo stars. The thin disk scale length, however, is not well-constrained because we are limited by the lack of coverage in $R$ and $|Z|$, particularly in the plane of the Galaxy. Future surveys such as APOGEE \citep{eis11} will be able to provide stricter constraints on both scale lengths.

\bibliographystyle{apj}
\bibliography{bib_alphas}

\end{document}